\documentclass[aps,prd,nofootinbib,twocolumn,superscriptaddress,preprintnumbers]{revtex4-1}

\usepackage{amssymb}
\usepackage{amsmath}
\usepackage{epsfig}
\usepackage{hyperref}
\usepackage{breakurl}
\usepackage{bm}
\usepackage{color}
\usepackage{tikz}
\usetikzlibrary{decorations.markings}

\makeatletter
\def\simgt{\mathrel{\lower2.5pt\vbox{\lineskip=0pt\baselineskip=0pt
           \hbox{$>$}\hbox{$\sim$}}}}
\def\simlt{\mathrel{\lower2.5pt\vbox{\lineskip=0pt\baselineskip=0pt
           \hbox{$<$}\hbox{$\sim$}}}}

\def\fig#1{Fig.~\ref{#1}}

\makeatother

\def\eqn#1{Eq.~\eqref{#1}}
\def\spa#1.#2{\left\langle#1\,#2\right\rangle}
\def\spb#1.#2{\left[#1\,#2\right]}
\def\sand#1.#2.#3{%
\left\langle#1{\vphantom1}\right|{#2}\left|#3\right]}%
\def\sandmp#1.#2.#3{%
\left\langle#1{\vphantom1}\right|{#2}\left|#3\right]}%
\def\sandpm#1.#2.#3{%
\left[#1{\vphantom1}\right|{#2}\left|#3\right\rangle}%
\def\sandmm#1.#2.#3{%
\left\langle#1{\vphantom1}\right|{#2}\left|#3\right\rangle}%
\def\sandpp#1.#2.#3{%
\left[#1{\vphantom1}\right|{#2}\left|#3\right]}%

\def\pp{\sigma}
\def\nn{\nonumber}
\def\Section#1{\vskip .05 cm \noindent {\it #1}}
\newcommand{\be}{\begin{equation}}
\newcommand{\ee}{\end{equation}}
\newcommand{\eq}[2]{\be\begin{aligned}#1 \label{#2}\end{aligned}\ee}

\newcommand{\Ref}[1]{Ref.~\cite{#1}}
\newcommand{\Fig}[1]{Fig.~\ref{#1}}
\newcommand{\Eq}[1]{Eq.~\eqref{#1}}

\def\topbotatom#1{\hbox{\hbox to 0pt{$#1\bot$\hss}$#1\top$}}

\begin{document}

\title{Scattering Amplitudes and the Conservative Hamiltonian \\ for Binary Systems at Third Post-Minkowskian Order}
\author{Zvi Bern}
\affiliation{
Mani L. Bhaumik Institute for Theoretical Physics,
University of California at Los Angeles,
Los Angeles, CA 90095, USA}
\author{Clifford Cheung}
\affiliation{Walter Burke Institute for Theoretical Physics,
    California Institute of Technology, Pasadena, CA 91125}
\author{Radu Roiban}
\affiliation{Institute for Gravitation and the Cosmos,
Pennsylvania State University,
University Park, PA 16802, USA}
\author{Chia-Hsien Shen}
\affiliation{
Mani L. Bhaumik Institute for Theoretical Physics,
University of California at Los Angeles,
Los Angeles, CA 90095, USA}
\author{ Mikhail P. Solon}
\affiliation{Walter Burke Institute for Theoretical Physics,
    California Institute of Technology, Pasadena, CA 91125}
\author{Mao Zeng}
\affiliation{Institute for Theoretical Physics, ETH Z\"urich, 8093 Z\"urich, Switzerland}

\begin{abstract}

We present the amplitude for classical scattering of gravitationally
interacting massive scalars at third post-Minkowskian order.  Our approach
harnesses powerful tools from the modern amplitudes program such as generalized
unitarity and the double-copy construction, which relates gravity
integrands to simpler gauge-theory expressions.  Adapting methods for
integration and matching from effective field theory, we 
extract the conservative Hamiltonian for compact spinless binaries at
third post-Minkowskian order.  The resulting Hamiltonian is in complete
agreement with corresponding terms in state-of-the-art expressions at
fourth post-Newtonian order as well as the probe limit at all orders in
velocity.   We also derive the scattering angle at third post-Minkowskian order.
\end{abstract}

\preprint{CALT-TH 2019-002 \hskip 12 cm UCLA/TEP/2019/101} 

\maketitle

\Section{Introduction.}
The recent discovery of gravitational waves at LIGO/Virgo~\cite{LIGO}
has launched an extraordinary new era in multi-messenger astronomy.
Given expected improvements in detector sensitivity, high-precision
theoretical predictions from general relativity will be crucial.
Existing theory benchmarks come from a variety of approaches (see also
\Ref{GravityReviews} and references therein), including the effective
one-body formalism \cite{EOB}, numerical relativity~\cite{NR}, the
self-force formalism~\cite{self_force}, and perturbative analysis
using post-Newtonian
(PN)~\cite{PN,Jaranowski:1997ky,DJSBlanchet3PN,JaranowskSchafer},
post-Minkowskian (PM)~\cite{PM, Damour:2016gwp, DamourTwoLoop}, and
effective field theory (EFT)~\cite{EFT} methods.

The past decade has also witnessed immense progress in the study of
scattering amplitudes, where understanding mathematical structures within
gauge theory and gravity has yielded new physical insights and
efficient methods for calculation.  In particular, the
Bern-Carrasco-Johansson (BCJ) color-kinematics duality and associated
double copy construction~\cite{BCJ} allow multiloop gravitational
amplitudes to be constructed from sums of products of gauge-theory
quantities. This has yielded a variety of new results in supergravity (see
\Ref{SimplifyMultiloopGrav} for recent results).  The BCJ construction
is intimately tied to the Kawai-Lewellen-Tye (KLT)
relations~\cite{KLT}, which relate tree amplitudes of closed and
open strings.

In this paper, we apply modern amplitude methods to derive the
classical scattering amplitude for two massive spinless particles at
${\cal{O}}(G^3)$ and to all orders in the velocity, {\it i.e.}~at the
third post-Minkowskian (3PM) order.  We use generalized
unitarity~\cite{GeneralizedUnitarity} to construct the corresponding
two-loop integrand from tree amplitudes of gravitons and massive
scalars, obtained straightforwardly from the double-copy
construction.  While the double copy introduces dilaton and antisymmetric
tensor degrees of freedom~\cite{Dilaton} which are absent in pure
Einstein gravity, we remove these unwanted
states efficiently by restricting the state sums in unitarity cuts to gravitons
alone.  As we will show, we can calculate in strictly
$D=4$ dimensions for the classical dynamics, where spinor
helicity variables~\cite{SpinorHelicity,AmplitudeReview} dramatically simplify the
required tree amplitudes.  The viability of working in $D=4$
offers optimism for extending our results to higher
orders.

Afterwards, we integrate the two-loop integrand via a procedure
adapted from EFT, in which energy integrals are evaluated in the
potential region via residues before performing spatial
integrations~\cite{2PM}.  Using EFT matching
\cite{RothsteinClassical,2PM} we then derive the 3PM conservative
Hamiltonian for compact spinless binaries. We show that the 4PN terms
in our Hamiltonian are, up to a coordinate transformation, physically
equivalent to corresponding terms in state-of-the-art results.
We also verify that our
result agrees in the probe limit with the Hamiltonian for a test body
orbiting a Schwarzschild black hole to 3PM order.  
Finally, we derive a compact expression for the 3PM scattering angle in terms of amplitude data.

\begin{figure}
\begin{center}
\includegraphics[scale=.31]{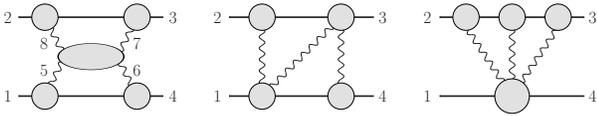}
\end{center}
\vskip -.5cm
\caption{Unitarity cuts needed for the classical scattering amplitude.
  The shaded ovals represent tree amplitudes while the exposed lines
  depict on-shell states.  The wiggly and straight lines denote
  gravitons and massive scalars, respectively.}
\label{OneLoopHelFigure}
\end{figure}

\Section{Double copy and unitarity.}
Dynamics at 3PM order is encoded in the two-loop scattering amplitude
for two massive, gravitationally interacting scalars.  Our calculation
begins with a construction of the corresponding two-loop integrand via
generalized unitarity.  Because we are interested in classical
scattering, we need not assemble the full quantum-mechanical integrand.
Rather, as emphasized in Refs.~\cite{RothsteinClassical, BjerrumGrav,
  2PM},  the classical potential only receives contributions with a single
on-shell matter line per loop and with no gravitons starting and ending on
the same matter line.  For this reason we focus solely on the
unitarity cuts shown in \fig{OneLoopHelFigure}.
 
We obtain the tree amplitudes in the unitarity cuts via two methods.
In the first approach, we work in general $D$ space-time dimensions.  Exploiting
color-kinematics duality~\cite{BCJ}, we derive
gravitational amplitudes straightforwardly 
from simpler gauge-theory amplitudes by
replacing color factors with corresponding kinematic factors.  For the
unitarity cuts of the classical limit of the two-loop scattering amplitude, the reference
momenta that complicate projection onto graviton physical states
can be eliminated, simplifying the
calculation~\cite{FutureDetailsLongPaper}. 
The primary purpose of our $D$-dimensional construction is to confirm explicitly the completeness of our second method, where we work in strictly $D=4$ so as to benefit from very simple expressions for
gauge-theory amplitudes in terms of spinor
helicity~\cite{SpinorHelicity} variables.  We then build the
two corresponding gravitational amplitudes via the KLT
relations~\cite{KLT}.  At two loops, both approaches are efficient, but
at higher loops, helicity amplitudes offer a much more compact
starting point.  

For concreteness, consider the first generalized unitarity cut in \fig{OneLoopHelFigure}, 
which we refer to as $C^\text{H-cut}$ and is comprised of 
products of four three-point and one four-point amplitudes.  Since four-point tree amplitudes are already very simple 
there is little computational advantage to imposing the on-shell conditions on matter lines.  Thus, we replace the
pairs of three-point amplitudes at the top and bottom of the cut with four-point amplitudes and then impose the matter 
cut conditions at the end.   
The resulting iterated two-particle cut is then
\begin{align}
C^{2,2} &= \sum_{\rm states}  M_4(2^s,-8, 7, 3^s)
 M_4(-5,6,-7,8)
\nn 
\\
& \hskip 1.75 cm 
\times M_4(1^s, 5, -6,4^s) ,
\label{Cut22}
\end{align}
where $M_4$ denotes the tree-level four-point amplitude for gravity
minimally coupled to two massive scalars denoted here by legs $1^s$, $2^s$, $3^s$, $4^s$.
In this cut, legs $1^s$, $4^s$ have mass $m_1$ while legs $2^s$, $3^s$ have mass $m_2$.
All momenta in each tree amplitude are taken to be outgoing.
The sum runs over graviton states for legs 5, 6, 7, 8, where the minus signs on the labels 
indicate reversed momenta.

The four-point gravity tree amplitudes needed in the cuts are obtained 
from gauge-theory ones via the field-theory limit of KLT relations~\cite{KLT}, 
\begin{align}
& M_4 (1,2,3,4)  = - i s_{12} A_4(1,2,3,4) \,
   A_4(1,2,4,3),
\label{KLT}
\end{align}
where the $A_4$ are tree-level color-ordered gauge-theory four-point
amplitudes and $s_{ij} = (p_i + p_j)^2$, working in mostly minus metric signature throughout.  Strictly speaking, the KLT
relations apply only to massless states.  However, they can be applied here
by interpreting the scalar masses, in the sense of dimensional
reduction, as extra-dimensional momentum components.  While we have
not included coupling constants, these are easily restored at the end
of the calculation by including an overall factor of $(8\pi G)^3$, where $G$ is Newton's constant.

In terms of the spinor-helicity conventions of Ref.~\cite{AmplitudeReview},
the independent tree-level gauge-theory amplitudes needed in \Eq{Cut22} are
\begin{align}
A_4(1^s,2^+,3^+,4^s) & = i\, \frac{ m_1^2 \spb{2}.{3} }
    { \spa{2}.{3} t_{12}}\, , \nonumber \\
A_4(1^s,2^+,3^-, 4^s) & =
    i\, \frac{\sandmp{3}.{1}.{2}^2}{t_{23}\, t_{12} } \, , \nonumber \\
A_4(1^-, 2^-, 3^+, 4^+) & =  i\, \frac{\spa{1}.{2}^4}{\spa{1}.{2}\spa{2}.{3} \spa{3}.{4} \spa{4}.{1}}  , \nonumber\\
A_4(1^-, 2^+, 3^-, 4^+) & =  i\, \frac{\spa{1}.{3}^4}{\spa{1}.{2}\spa{2}.{3} \spa{3}.{4} \spa{4}.{1}}  ,
\end{align}
where $t_{ij} = 2 p_i \cdot p_j$ and the $\pm$ denote gluon helicities.

The dilaton and antisymmetric tensor states are removed from unitarity cuts
by correlating the gluon helicities on both sides of the double copy. The unwanted states correspond to 
one gluon in the double copy of positive helicity and the other of negative helicity. 
An internal graviton state is obtained by taking the corresponding gluons in 
the KLT formula  in \Eq{KLT} to be of the same helicity.

Using spinor evaluation techniques, it
is straightforward to obtain a compact expression for the iterated
two-particle cut in \Eq{Cut22} ({\it e.g.}~see Ref.~\cite{TwoLoopN8}). 
Imposing cuts on the matter lines, as indicated in the first unitarity cut of \Fig{OneLoopHelFigure},  
further simplifies it and gives $C^\text{H-cut}$. We find
\begin{align}
&C^{\textrm{H-cut}}  =
  2i  \biggl[\frac{1}{(p_5 - p_8)^2} + \frac{1}{(p_5 + p_7)^2}
          \biggl] 
\nn \\
& \quad \times  \biggl[  s_{23}^2 m_1^4 m_2^4 + \frac{1}{s_{23}^6} \sum_{i=1,2} \Bigl(
  {\cal E}_i^4 + {\cal O}_i^4 + 6 {\cal O}_i^2 {\cal E}_i^2 \Bigr) \biggr] 
 ,
\label{HCut}
\end{align}
where we have defined
\begin{align}
{\cal E}_1^2 & = \frac{1}{4} s_{23}^2 (t_{1 8} t_{25} - t_{12} t_{5 8})^2 
,  \hskip .5 cm 
{\cal O}_1^2  = {\cal E}_1^2 - m_1^2 m_2^2 s_{23}^2 t_{58}^2 , \hskip .2 cm \nn\\
{\cal E}_2^2 & =  \frac{1}{4}  s_{23}^2  (t_{17} t_{25} - t_{12} t_{57} -
                                s_{23}(t_{17} + t_{57} ))^2 , \nn \\
{\cal O}_2^2 &= {\cal E}_2^2  -  m_1^2 m_2^2 s_{23}^2 t_{5 7}^2.
\end{align}
The simplicity of this expression is a reflection of the double-copy
structure: the same building blocks appear in the simpler corresponding
gauge-theory cut.

The spurious double-pole in $s_{23}$ can be explicitly cancelled by
adding terms proportional to the Gram determinant formed from the five
independent momenta at two loops which vanishes in $D=4$.  In fact,
the expression derived from the $D$-dimensional approach is automatically
free of such spurious singularities.  While these Gram determinants
contribute quantum mechanically, we have checked explicitly that they
vanish in the classical limit.  This is not accidental---such terms
are of the wrong form to generate the required $\log (s_{23})$ needed to
contribute to the classical 3PM amplitude (see \Ref{FutureDetailsLongPaper} for details).

\begin{figure}
\begin{center}
\includegraphics[scale=.27]{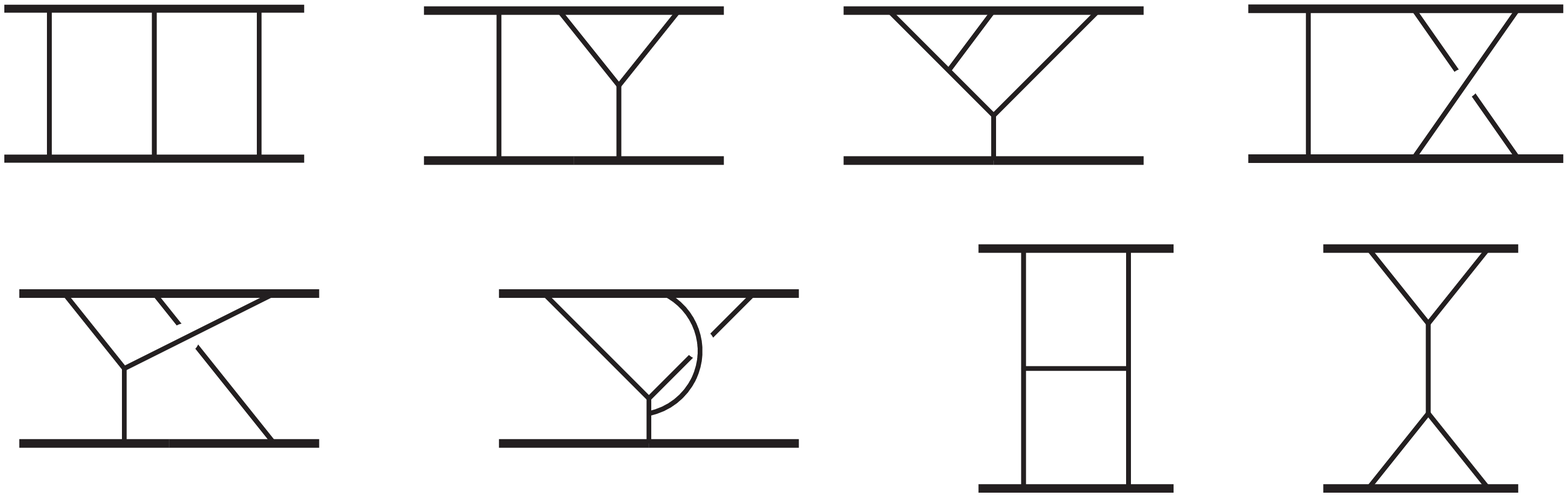}
\end{center}
\vskip -.45cm
\caption{The eight independent diagrams showing the propagator structure of integrals 
from which the classical contributions are extracted.}
\label{MasterDiagsFigure}
\end{figure}

The remaining two independent generalized unitarity cuts in
\fig{MasterDiagsFigure} are more complicated because they require
five-point tree amplitudes with two massive scalar legs.  The
four-dimensional input gauge-theory amplitudes are simple to compute
using modern methods ({\it e.g.}~see Ref.~\cite{KhozeMassive}). For our $D$
construction we obtain a BCJ representation, allowing us to express
the gravity cuts directly in terms of local diagrams. The particular
representation was chosen such that we can ignore the reference momenta 
when projecting the internal states into gravitons. Further details will be given elsewhere~\cite{FutureDetailsLongPaper}.

To facilitate integration, we merge the cuts into a single integrand
whose cuts match those in \fig{OneLoopHelFigure}. This is achieved using  an ansatz in terms of eight independent diagrams with
only cubic vertices displayed in \fig{MasterDiagsFigure}.  The
diagrammatic numerators are polynomials of the appropriate dimension
exhibiting the symmetries of the corresponding diagram.  Their coefficients are then
fixed via the method of maximal cuts~\cite{MaximalCutMethod}, whereby 
cuts of the integrand are constrained to match the known ones.  This approach is sufficient
for the two-loop problem.  

\Section{Integration.}
Our method of integration follows
Ref.~\cite{2PM}.  For convenience, we give a short summary here,
leaving details to Ref.~\cite{FutureDetailsLongPaper}.  Terms in the integrand take the form,
\eq{
{\cal I} &= \frac{\textrm{numerator}}{\textrm{graviton propagators}}
 \times
 \prod_{i} \frac{1}{\omega_i^2 -  \bm k_i^2 -m_i^2} ,
}{}
where $i$ labels each matter line, which has energy $\omega_i$, spatial momentum $ \bm k_i$, and mass $m_i$.   
The matter propagators can be factored into particle and antiparticle poles, ${\omega_i \pm  \sqrt{\bm k_i^2 + m_i^2}}$.
We then express the integrand as ${\cal I} = {\cal N} \times \prod_{i} \frac{1}{z_i}$, {\it i.e.}~in terms of the particle poles $z_i  = \omega_i -  \sqrt{ \bm k_i^2 + m_i^2}$ and an effective numerator ${\cal N}$ which absorbs the rest of the integrand.

Following the procedure outlined in \Ref{2PM}, we first evaluate the energy integrals.
At two loops, {\it i.e.}~3PM order, we integrate over two independent combinations of energies, $\omega$ and $\omega'$, in the potential region.  As we will prove in detail in \Ref{FutureDetailsLongPaper}, the result is
\eq{
\widetilde {\cal I} = \int\! \frac{d\omega}{2\pi} \, \frac{d\omega'}{2\pi} \, {\cal I}(\omega,\omega') &= \sum_{(i,j)} S_{ij}  \underset{\;\,
 \omega_{ij}, \omega'_{ij}}{\textrm{Res}} {\cal I}(\omega,\omega'),
}{}
where the sum runs over distinct pairings $(i,j)$ of matter poles and $z_i = z_j=0$ when $(\omega, \omega') = (\omega_{ij} , \omega_{ij}')$.
Here $S_{ij}$ is a calculable symmetry factor whose sign and magnitude depend on the topology of the cut graph.  Note that the residue for an $(i,j)$ pairing  will vanish if there are no values of $\omega$ and $\omega'$ for which $z_i=z_j=0$.  

The resulting quantity $\widetilde {\cal I}$ depends on two independent spatial loop momenta.  To integrate over them
we employ dimensional regularization to deal with ultraviolet divergences stemming from the renormalization of delta function contact interactions which do not contribute classically.
Due to the localization on energy residues, $\widetilde {\cal I}$ is a complicated, non-polynomial function of three-dimensional invariants involving square roots.  Nevertheless, we can series expand $\widetilde {\cal I}$ in large $m_{1,2}$, yielding polynomials of kinematic invariants which we can integrate at each order.  After expanding, nearly all the spatial integrals are simple bubbles for which there are known analytic expressions~\cite{Smirnov:2012gma}.  The remaining integrals are evaluated via integration-by-parts identities~\cite{Chetyrkin:1981qh}.

For diagrams free from infrared (IR) singularities generated by iterations of lower-loop graviton exchanges, we have checked that our integrated results accord with several standard methods in the Feynman integral literature, including the Mellin-Barnes representation \cite{MellinBarnes, Smirnov:2012gma}, numerical integration via sector decomposition \cite{Smirnov:2015mct}, and differential equations \cite{DifferentialEquations} derived through integration-by-parts reduction \cite{Chetyrkin:1981qh, IBPReduction}. Since the classical contribution comes from certain residues on matter poles, the system of differential equations omits integrals without support on such residues.

\Section{Amplitude and potential.}
We apply the integration procedure outlined above order by order in the large-mass expansion, {\it i.e.}~in powers of velocity. 
Combining an explicit evaluation of the 3PM amplitude up to 7PN order with information on the pole structure of individual integrals and
exact, manifestly relativistic analytic results for certain graph topologies, we conjecture a full, all orders in velocity expression for the 3PM amplitude (whose uniqueness will be discussed in Ref.~\cite{FutureDetailsLongPaper}):
\begin{widetext}
\eq{
{\cal M}_3 =  & \frac{\pi  G^3 \nu ^2 m^4  \log {{\bm q}^2}}{6 \gamma ^2 \xi }  \bigg[   3-6\nu  + 206 \nu  \pp -54\pp^2 + 108 \nu \pp^2  + 4 \nu \pp^3  - {48 \nu \left(3+12 \pp^2-  4 \pp^4  \right) {\rm arcsinh} \sqrt{\pp-1 \over 2} \over \sqrt{ \pp^2 - 1}} \bigg. \\
& 
- {18 \nu \gamma \left( 1 - 2 \pp^2  \right) \left( 1 - 5 \pp^2 \right)  \over  \left( 1+ \gamma \right) \left(1 +  \pp \right) }\bigg] + {8\pi^3 G^3 \nu^4 m^6  \over \gamma^4 \xi } \bigg[ {3 \gamma  \left( 1 - 2 \pp^2  \right) \left( 1 - 5 \pp^2 \right)  } F_1  -{32 m^2 \nu^2 \left( 1 - 2 \pp^2  \right)^3}  F_2 \bigg]   ,
}{eq:A_3PM}
\end{widetext}
where the log scale dependence is absorbed into a delta-function ultraviolet counterterm.
Here we use center-of-mass coordinates where the incoming and outgoing  particle momenta are $\pm \bm p$ and $\pm (\bm p-\bm q)$, respectively.  We emphasize that  ${\cal M}_3$ includes the nonrelativistic normalization factor, $1/4 E_1 E_2$, where $E_{1,2} = \sqrt{\bm p^2 + m_{1,2}^2}$.  We  also define the total mass $m= m_1 +m_2$, the symmetric mass ratio $\nu = m_1 m_2 /m^2$, the total energy $E = E_1 +E_2$, the symmetric energy ratio $\xi = E_1 E_2 / E^2$, the energy-mass ratio $\gamma = E/m$, and the relativistic kinematic invariant $\pp = {p_1 \cdot p_2  \over m_1 m_2}$.  Note that the arcsinh factor is actually proportional to the sum of particle rapidities, ${\rm arctanh}\, |{{\bm p}}|/E_{1,2}$.

 \Eq{eq:A_3PM} only includes $\bm q$-dependent terms which persist in the classical limit.  In particular, the $\log \bm q^2$ term ultimately feeds into the conservative Hamiltonian through the Fourier transform $\left[ \log {\bm q}^2 \right]_{\rm FT} =  -{1 \over 2\pi |{\bm r}|^3} $.   Meanwhile the remaining IR-divergent contributions, parameterized by $F_1 = \int_{\bm k_1} \frac{1}{X_1^2 Y_1 X_2}$ and $F_2 = \int _{\bm k_1,\bm k_2} \frac{1}{X_1^2 Y_1 X_2^2 Y_2 X_3^2}$
in the notation described in Eq.(12) of Ref.~\cite{2PM}, will cancel in the EFT matching.  

The Hamiltonian is extracted from the amplitude via EFT methods
developed in Refs.~\cite{RothsteinClassical,2PM,FutureEFT} (see Ref.~\cite{DamourTwoLoop} for another approach).  Consider massive spinless
particles interacting via the center-of-mass Hamiltonian 
\eq{
H(\bm p, \bm r) &= \sqrt{\bm p^2 + m_1^2}+ \sqrt{\bm p^2 + m_2^2} +V(\bm p, \bm r), \\
V(\bm p, \bm r) & = \sum_{i=1}^\infty c_i(\bm p^2)  \left( \frac{G}{|\bm r|} \right)^i  ,
}{Hamiltonian3PM}
where $\bm r$ is the distance vector between particles and $i$ labels PM orders. The above Hamiltonian is in a gauge in which terms involving $\bm p \cdot \bm r$ or time derivatives of $\bm p$ are absent.  We 
then compute the scattering amplitude of massive scalars,
${\cal M}^{(\textrm{EFT})} = \sum_{i=1}^\infty {\cal M}^{(\textrm{EFT})}_{i}$, where ${\cal M}^{(\textrm{EFT})}_{3}$ comes from diagrams with two or
fewer loops that depend on $c_1$, $c_2$, and $c_3$.  In \Ref{2PM}, the coefficients
$c_1$ and $c_2$ were extracted analytically to all orders in velocity.  Inserting these into
${\cal M}_{3}^{(\textrm{EFT})}$ effectively implements the
subtraction of iterated contributions.  By equating
${\cal M}_{3}^{(\textrm{EFT})} = {\cal M}_{3}$, we solve
for the 3PM coefficient $c_3$.

The main result of the present work is the 3PM potential, encapsulated in the coefficients
\begin{widetext}
\eq{
&c_1 = \frac{\nu ^2 m^2}{\gamma ^2 \xi } \left(1-2 \sigma ^2\right), 
\qquad
c_2 =\frac{\nu ^2 m^3}{\gamma ^2 \xi } \left[ \frac{3}{4} \left(1-5 \sigma
   ^2\right)-\frac{4 \nu  \sigma  \left(1-2 \sigma
   ^2\right)}{\gamma  \xi } -
\frac{\nu ^2 (1-\xi) \left(1-2 \sigma ^2\right)^2}{2 \gamma ^3
   \xi ^2}
\right],
\\
&c_3 = 
\frac{\nu ^2 m^4}{\gamma ^2 \xi } \Biggl[   \frac{1}{12}{\left(3-6\nu  + 206 \nu  \pp -54\pp^2 + 108 \nu \pp^2  + 4 \nu \pp^3\right)}
-\frac{4 \nu  \left(3+12 \pp^2-  4 \pp^4  \right) {\rm arcsinh} \sqrt{\frac{\sigma
   -1}{2} }}{\sqrt{\sigma ^2-1}}
 \\
& \hskip 1.8 cm  \null
   -\frac{3 \nu \gamma   \left(1-2 \sigma ^2\right) \left(1-5
   \sigma ^2\right)}{2(1+\gamma)(1+\pp) }
-\frac{3 \nu  \sigma  \left(7-20 \sigma ^2\right)}{2 \gamma  \xi
   }
   -\frac{\nu ^2  \left(3+8\gamma - 3\xi
   -15\pp^2 -   80 \gamma 
   \sigma ^2 +15 \xi  \sigma ^2 \right)\left(1-2 \sigma ^2\right)}{4 \gamma ^3 \xi ^2}
   \\
& \hskip 1.8 cm  \null
   +\frac{2 \nu ^3 (3-4 \xi ) \sigma  \left(1-2 \sigma ^2\right)^2}{\gamma ^4 \xi ^3}          
   +\frac{\nu ^4 (1-2 \xi) \left(1-2 \sigma ^2\right)^3}{2 \gamma
   ^6 \xi ^4} \,
    \Biggr] ,
}{eq:V_3PM}
\end{widetext}
where for convenience, the expressions for $c_1$ and $c_2$ in \Ref{2PM} are reproduced here with slightly different normalization and in our current notation. As emphasized in \Ref{2PM}, the cancellation of IR divergences between ${\cal M}_{3}^{(\textrm{EFT})}$ and ${\cal M}_{3}$ depends critically on $c_1$ and $c_2$ and thus provides a nontrivial check of our calculation.

\Section{Consistency checks.}
Our results pass several nontrivial albeit overlapping consistency checks (see Ref.~\cite{FutureDetailsLongPaper} for details). First and foremost, we have verified that the 4PN terms in our Hamiltonian are equivalent to known results up to a canonical coordinate transformation,
\eq{
&(\bm {r} , \bm p)  \rightarrow (\bm {R},\bm {P}) = ( A \, \bm r   + B\,  \bm p  , C \, \bm p  +  D \, \bm r ) \\
&A  = 1 - \frac{Gm \nu}{2|\bm r|} + \cdots, \quad \! B = \frac{G (1-2/\nu)}{4m |\bm r|} \bm p\cdot \bm r+ \cdots,\\
&C  = 1 + \frac{Gm \nu}{2| \bm r|} + \cdots,  \quad \! D = -\frac{G m  \nu}{2 | \bm r|^3} \bm p\cdot \bm r+ \cdots, 
}{}
with ellipses denoting higher order corrections entering as a power series in $G/| \bm r|$, $\bm p^2$, and $(\bm p\cdot \bm r )^2/ \bm r^2$ (for past treatments, see \Ref{BjerrumBohr:2002kt, Holstein:2008sx}).   To derive this coordinate transformation we generate an ansatz for
$A,B,C,D$ and constrain it to preserve the Poisson brackets, {\it i.e.}~$\{ \bm r,  \bm p \} =  \{ \bm R, \bm P \}  = \bm 1$ with all other brackets vanishing,
in the spirit of Ref.~\cite{DamourPoincare}.  
We verify that within this space of canonical transformations exists a subspace which maps our Hamiltonian in \Eq{eq:V_3PM} to 
the one in the literature, {\it e.g.}~as summarized in Eq.(8.41) of Ref.~\cite{JaranowskSchafer}, up to the intersection of 3PM and 4PN accuracy.

Second, applying the methods of Ref.~\cite{2PM} we have checked that the full-theory amplitude ${\cal M}_{3}$ in \Eq{eq:A_3PM} is identical to the amplitude ${\cal M}_{3}^{(\textrm{EFT})}$ computed from the conservative Hamiltonian in Ref.~\cite{JaranowskSchafer} up to 4PN accuracy.

Third, we have extracted from our Hamiltonian the coordinate invariant energy of a circular orbit as a function of the period.  Working at 2PN order---the highest order subsumed by 3PM which is relevant to a virialized system---we agree with known results~\cite{DJSBlanchet3PN}.

Fourth, we have extracted from our Hamiltonian the 3PM-accurate classical scattering angle in the center-of-mass frame, given by
\begin{align}
2\pi \chi =  {d_1 \over  J } + {d_2  \over J^2  }  + \frac{1}{J^3}
 \biggl( \vphantom{\frac{1}{\pi^2}} {  -4  d_3     }   
 + {  d_1 d_2 \over \pi^2    }  
-{ d_1^3 \over 48 \pi^2  }  \biggr),
\label{angle}
\end{align}
where $J = b |{\bm p} | $ is the angular momentum, $b$ is the impact parameter, and we have defined $d_1 = m \gamma \xi {\bm q}^2 {\cal M}_{1}'/|{\bf p}| $,
 $d_2 =m \gamma \xi |\bm q| {\cal M}_{2}' $, 
and $d_3 =  m \gamma \xi |{\bf p}| {\cal M}_{3}' / \log {\bm q}^2 $.
The primed quantities denote 
the IR-finite parts of the nonrelativistically normalized amplitudes that enter the Hamiltonian coefficients
as defined here and in \Ref{2PM}, so  
\begin{align}
{\cal M}_{1}' & = -\frac{4 \pi G\nu^2 m^2 }{\gamma^2 \xi \bm q^2 }(1-2 \sigma^2), \nonumber \\
{\cal M}_{2}' & = -\frac{3 \pi^2 G^2 \nu^2 m^3 }{2 \gamma^2 \xi |\bm q|}(1-5 \sigma^2),
\end{align}
and
${\cal M}_{3}'$ is the $\log {\bm q}^2$ term in  \Eq{eq:A_3PM}.  Truncated to 4PN order, \eqn{angle} agrees with known results~\cite{Bini:2017wfr}.

Last but not least, in the probe limit $m_1 \ll  m_2$, our result exactly coincides with the Hamiltonian for a point particle in a Schwarzschild background to ${\cal O}(G^3)$ and all orders in velocity, {\it e.g.}~as given in Eq.(8) of Ref.~\cite{WS93}.

\Section{Conclusions.}
We have presented the 3PM amplitude for classical scattering of
gravitationally interacting massive spinless particles.  From this
amplitude we have extracted the corresponding conservative Hamiltonian
for binary dynamics to 3PM order.  

The 3PM Hamiltonian in Eqs.~\eqref{Hamiltonian3PM} and~\eqref{eq:V_3PM} will be employed in a forthcoming paper~\cite{BuonannoEnergy}
to compute approximants for the  binding energy of binary systems moving on circular orbits
and assess their accuracy against numerical-relativity predictions. This is relevant for understanding
the usefulness of PM calculations when building accurate waveform models for LIGO/Virgo data analysis.

Our paper leaves many avenues for future work, {\it
  e.g.}~including obtaining higher orders in the PM expansion, incorporating
spin~\cite{spin}, radiation~\cite{radiation}, and finite-size effects, as well as connecting to other recent amplitude
approaches~\cite{Dilaton,OtherAmpitude} and the effective one-body formalism~\cite{EOB,
 Damour:2014afa,  Damour:2016gwp, DamourTwoLoop}.

The simplicity of the 3PM amplitude in
\Eq{eq:A_3PM} and potential in \Eq{eq:V_3PM} bodes well for future
progress. Moreover, since the amplitude and EFT methods employed in this paper are far from
exhausted, we believe that the results we have reported mark
only the beginning.

\Section{Acknowledgements.}
We thank Alessandra Buonanno, Thibault Damour, Michael Enciso, David
Kosower, Andr\'es Luna, Aneesh Manohar, Smadar Naoz, Julio Parra-Martinez, Rafael
Porto, Jan Steinhoff, George Sterman, Gabriele Veneziano, Justin Vines, and Mark Wise for helpful discussions, including comments on the manuscript. 
 In addition, we especially thank Ira Rothstein for his many insightful
comments throughout this project.  Z.B. is supported by the U.S. Department of Energy (DOE) 
under Award Number DE-SC0009937.  C.C.~is supported by the DOE under grant no.~DE-SC0011632.  R.R.~is
supported by the U.S. Department of Energy (DOE) under grant
no.~DE-SC0013699.  C.H.S.~is supported by the Mani L. Bhaumik Institute
for Theoretical Physics. M.P.S.~is supported by the DOE under grant
no.~DE-SC0011632 and the McCone Fellowship at the Walter Burke
Institute.  M.Z.~is supported by the Swiss National Science Foundation under
contract SNF200021 179016 and the European Commission through the ERC grant pertQCD.


\end{document}